\documentclass[conference]{IEEEtran}
\IEEEoverridecommandlockouts

\usepackage{cite}
\usepackage{amsmath,amssymb,amsfonts}
\usepackage{algorithmic}
\usepackage{graphicx}
\usepackage{textcomp}
\usepackage[table]{xcolor}
\usepackage{booktabs}
\usepackage{tabularx} 
\usepackage{multirow}
\usepackage{algorithm}
\usepackage{tikz}
\usepackage{pgfplots}
\usepackage{acronym}
\pgfplotsset{compat=1.17}

\newcolumntype{Y}{>{\raggedright\arraybackslash\hsize=1.5\hsize}X}
\newcolumntype{Z}{>{\raggedright\arraybackslash\hsize=0.5\hsize}X}

\newacro{AE}{Average Error}
\newacro{AI}{Artificial Intelligence}
\newacro{AP}{Access Point}
\newacro{BSS}{Basic Service Set}
\newacro{BSSID}{Basic Service Set Identifier}
\newacro{HT}{High Throughput}
\newacro{IE}{Information Element}
\newacro{IFAT}{Inter-Probe Frame Arrival Time}
\newacro{IRM}{Identifiable Random MAC Address}
\newacro{KDE}{Key Data Encapsulation}
\newacro{LDPC}{Low Density Parity Check}
\newacro{LSTM}{Long Short-Term Memory}
\newacro{MAC}{Medium Access Control}
\newacro{ML}{Machine Learning}
\newacro{MSRSSI}{Multi-Simulated Received Signal Strength Indication}
\newacro{NB}{Naive Bayes}
\newacro{NIC}{Network Interface Controller}
\newacro{NP}{Noisy Point}
\newacro{OS}{Operating System}
\newacro{OSI}{Open Systems Interconnection}
\newacro{OUI}{Organizationally Unique Identifier}
\newacro{PHY}{Physical Layer}
\newacro{RF}{Random Forest}
\newacro{RI}{Rand Index}
\newacro{RSSI}{Received Signal Strength Indication}
\newacro{SC}{Silhouette Coefficient}
\newacro{SRSSI}{Simulated Received Signal Strength Indication}
\newacro{SSID}{Service Set Identifier}
\newacro{STA}{Station}
\newacro{STBC}{Space-Time Block Coding}
\newacro{SVM}{Support Vector Machine}
\newacro{VHT}{Very High Throughput}
\newacro{VM}{V-Measure}
\newacro{EHT}{Extremely High Throughput}
\newacro{HE}{High Efficiency}
\newacro{LNSM}{Log-Normal Shadowing Model}
\newacro{SN}{Sequence Number}

\def\BibTeX{{\rm B\kern-.05em{\sc i\kern-.025em b}\kern-.08em
    T\kern-.1667em\lower.7ex\hbox{E}\kern-.125emX}}
\begin{document}

\title{Can Machine Learning Break Wi-Fi Privacy?\\A Study on MAC Address Randomization}

\author{
  \IEEEauthorblockN{
    Marta Puig$^{\star}$, Costas Michaelides$^{\star}$, Lucia Pintor$^{\dagger }$, Boris Bellalta$^{\star}$, and Francesc Wilhelmi$^{\star}$
  }
  \IEEEauthorblockA{
    $^{\star}$\textit{Universitat Pompeu Fabra}, Barcelona, Spain\\
    $^{\dagger}$\textit{University of Cagliari}, Cagliari, Italy\\
  }
  \IEEEauthorblockN{\thanks{Corresponding author: \emph{martazheng.puig@upf.edu}.}}
}

\maketitle

\begin{abstract}
\ac{MAC} address randomization has been widely adopted during the IEEE 802.11 network discovery phase as a countermeasure against passive tracking. This paper exposes vulnerabilities in these privacy protocols by demonstrating that devices remain identifiable using \ac{ML}-based fingerprinting. To study the potential tracking capabilities of a passive attacker, we evaluate different eavesdropping scenarios and configurations. To this end, we extract unencrypted hardware specifications from Probe Frames, which we combine with the \ac{IFAT} and \ac{SRSSI} signals. A core contribution of this paper is the bitwise decomposition of the \ac{HT} capabilities information field, which improves device identification accuracy. We evaluate this de-randomization approach using three unsupervised clustering algorithms (K-Means, DBSCAN, and OPTICS) across a dataset of 22 devices from six manufacturers. Our results show that DBSCAN, when using decomposed \ac{HT} capabilities information and three \ac{SRSSI} measurements, achieves a global accuracy up to 89.6\%. This suggests that the existing MAC randomization solutions are insufficient and underscores the need for enhancing privacy within Wi-Fi standardization.
\end{abstract}

\begin{IEEEkeywords}
IEEE 802.11, MAC address randomization, privacy, machine learning, Wi-Fi
\end{IEEEkeywords}

\acresetall

\section{Introduction}

Wi-Fi clients, or \acp{STA}, discover available networks using two methods: passive scanning, where they listen for Beacon frames, or active scanning, which involves transmitting Probe Request frames and receiving Probe Response frames from nearby \acp{AP}. A critical vulnerability of active discovery is that its management frames are transmitted entirely unencrypted, making them a primary target for passive eavesdropping and device tracking~\cite{henry2024wifi7}. 


To mitigate third-party tracking, vendors such as Apple (starting from iOS 8) and Microsoft (starting from Windows 10) introduced their own \ac{MAC} address randomization mechanisms before formal standardization. However, while this uncoordinated randomization improved privacy, it introduced new risks that broke basic 802.11 operations. Finally, the IEEE formally addressed \ac{MAC} address randomization in the 802.11aq amendment for Probe Requests during the pre-association stage~\cite{henry2024wifi7}. Despite randomization, Probe Requests still contain static header fields that can be used to track a given user. Moreover, temporal information can be extracted from \ac{IFAT} and spatial information can be derived from the \ac{RSSI}, which further contribute to passively fingerprinting users even when the \ac{MAC} address is randomized~\cite{Vanhoef2016}.


\begin{table*}[t]
\caption{Representative prior art on \ac{MAC} de-randomization. Feature columns: SN = \acf{SN}, IE = \acf{IE}, T = Timing, R = \ac{RSSI}.}
\label{tab:comparison_methods}
\centering
\footnotesize
\renewcommand{\arraystretch}{1.1}
\setlength{\tabcolsep}{4pt}

\begin{tabularx}{\textwidth}{@{}
l
c c c c
Y
Z
>{\raggedright\arraybackslash}p{2.8cm}
@{}}
    \toprule
    \textbf{Work} & \textbf{SN} & \textbf{IE} & \textbf{T} & \textbf{R} & \textbf{Environment \& Dataset} & \textbf{Algorithm} & \textbf{Result} \\
    \midrule
    Freudiger~\cite{Freudiger2015}
        & \checkmark & \checkmark & \texttimes & \texttimes
        & Laboratory (3 channels). Labeled (True): Specific devices with known \ac{OS}.
        & Linear & Proven \ac{SN} leak \\
    \addlinespace

    Matte et al.~\cite{Matte2016}
        & \texttimes & \texttimes & \checkmark & \texttimes
        & Laboratory (1 channel). Labeled (Simulated): 120k frames; simulated randomization.
        & D2 & 77.2\% Accuracy \\
    \addlinespace

    Praharenka et al.~\cite{Praharenka2021}
        & \texttimes & \texttimes & \checkmark & \texttimes
        & Isolated (3 channels). Labeled (True): Manual labeling in a controlled environment.
        & \ac{NB}/RF/\ac{SVM} & 94.5\% / 93.2\% / 90.4\% Accuracy \\
    \addlinespace

    Pintor et al.~\cite{Pintor2022,Pintor2024}
        & \texttimes & \checkmark & \texttimes & \texttimes
        & Isolated (3 channels). Labeled (True): Single-device captures merged into 12 scenarios.
        & DBSCAN & 7.5\% Average Error \\
    \addlinespace

    Cifuentes-Urtubey et al.~\cite{CifuentesUrtubey2024}
        & \checkmark & \checkmark & \texttimes & \texttimes
        & Conference venue (1 channel). Unlabeled: Real-world MobiCom 2023 traffic.
        & Heuristics & 45\% reduction in \ac{MAC} IDs \\
    \addlinespace

    P\'{e}rez-Hern\'{a}ndez et al.~\cite{PerezHernandez2024}
        & \texttimes & \checkmark & \texttimes & \checkmark
        & Building (1 channel). Unlabeled: Ground truth via manual counting.
        & K-Means & 98\% Counting Accuracy \\
    \addlinespace

    \textbf{This work}
        & \texttimes & \checkmark & \checkmark & \checkmark
        & \textbf{Isolated (1 channel). Labeled (True): Single-device captures merged into 3 scenarios.}
        & \textbf{DBSCAN} & \textbf{89.6\% Global Accuracy} \\
    \bottomrule
\end{tabularx}
\end{table*}

Prior fingerprinting methodologies generally exploit individual feature categories, including protocol header information, temporal behavior, and spatial signal characteristics, as summarized in Table~\ref{tab:comparison_methods}. While some previous works relied on the \acf{SN}, we omit it from our work because it no longer increments predictably across different bursts. To explore the combined effect of the remaining features, the present work builds upon the dataset presented in~\cite{Pintor2022,Pintor2024}. Since the original dataset lacks environmental context, \ac{RSSI} values are simulated here. In particular, we propose a framework that integrates \acp{IE}, \ac{IFAT}, and \ac{SRSSI} features to assess the robustness of \ac{MAC} address randomization against unsupervised \ac{ML} clustering. In a real-world eavesdropping scenario, an attacker aims to determine which randomized \acp{MAC} belong to the same device. Therefore, we use unsupervised clustering to group Probe Requests based on the information they carry, without needing any prior knowledge of the network. The evaluation was conducted on 22 heterogeneous devices from six manufacturers under four feature configurations and four device density levels (5, 10, 15, and 22 devices). 

Overall, we show that current \ac{MAC} randomization solutions are not enough, highlighting the need for further improvements to Wi-Fi standards. Specifically:


\begin{itemize}
    \item We show that the bitwise decomposition of \ac{HT} capabilities information into subfields leads to a higher separation among devices, making it easier for the attacker to track them.
    \item We show that the \ac{IFAT} feature introduces variance and degrades clustering performance across three unsupervised algorithms (K-Means, DBSCAN, and OPTICS).

    \item We show that having access to spatial information (simulated in this paper) improves de-randomization accuracy with respect to header-only features.
\end{itemize}

The remainder of the paper is structured as follows: Section~\ref{sec:methodology} describes the proposed methodology. 
Section~\ref{sec:experimental_setup} details the experimental setup.
Section~\ref{sec:results} presents the results and discussion. Section~\ref{sec:conclusion} concludes the paper.

\section{Proposed Methodology} \label{sec:methodology}
The proposed methodology consists of three blocks: Data Description and Preparation (Section~\ref{sec:data_description}), Clustering Algorithms (Section~\ref{sec:clustering_algorithms}), and Evaluation Metrics (Section~\ref{sec:evaluation_metrics}).

\subsection{Data Description and Preparation} \label{sec:data_description}
As previously mentioned, the dataset provided by \cite{Pintor2022} was utilized for this research. This dataset consists of 20-minute captures from 22 different devices under six distinct operational modes. Among all possible configurations, only captures in “Mode S” (screen inactive, Wi-Fi enabled, power-saving disabled) were considered. Analyzing the dataset more deeply, we observed a significant variance in the number of frames transmitted. Despite all devices being captured under the same conditions, certain devices broadcast considerably more frames within those 20 minutes, ranging from 5 frames (Device U) to over 1,200 frames (Device C). 

Table~\ref{tab:phy_standards} lists all the devices, identified by a letter, alongside the number of transmitted Probe Frames, observed \ac{PHY}, standard mode, and the hexadecimal \ac{HT} capability information. It should be noted that these \ac{PHY} types are the protocol functionalities advertised by the devices during the network discovery phase, and do not necessarily represent their overall hardware capabilities. Moreover, some devices exhibited two distinct \ac{HT} capability information values because they used different \ac{PHY} configurations during the capture. Consequently, there are two values in the Probe Frames count, one for each configuration.

\begin{table}[!t]
    \centering
    \caption{Devices with the number of transmitted Probe Frames and observed Wi-Fi \ac{PHY} types}
    \label{tab:phy_standards}
    \footnotesize 
    \renewcommand{\arraystretch}{1.1} 
    \setlength{\tabcolsep}{3pt} 

    \begin{tabularx}{\linewidth}{@{} c X c c c c @{}}
        \toprule
        \textbf{ID} & \textbf{Model} & 
        \begin{tabular}{@{}c@{}}\textbf{Number of} \\ \textbf{Probe Requests}\end{tabular} & 
        \textbf{\ac{PHY}} & \textbf{Standard} & 
        \begin{tabular}{@{}l@{}}\textbf{\ac{HT} Capabilities} \\ \textbf{Information}\end{tabular} \\
        \midrule
        B & Xiaomi Redmi 4 & 129 & 4 \& 6 & b/g & NaN \\
        E & Xiaomi Mi A2 Lite & 115 & 4 \& 6 & b/g & NaN \\
        J & Xiaomi Redmi 5 Plus & 123, 42 & 4 \& 6 & b/g & \texttt{0x016e}, NaN \\
        \midrule
        A & Samsung Galaxy M31 & 45 & 4 & b & \texttt{0x012d} \\
        C & Samsung Galaxy S4 & 1236 & 4 & b & \texttt{0x102d} \\
        D & Huawei P8 Lite & 6 & 4 & b & \texttt{0x102c} \\
        G & Huawei P20 & 617 & 4 & b & \texttt{0x0121} \\
        H & Samsung Gal. S6 Edge+ & 12 & 4 & b & \texttt{0x1163} \\
        I & Samsung Galaxy S7 & 28 & 4 & b & \texttt{0x1163} \\
        K & Samsung Galaxy J6 & 34 & 4 & b & \texttt{0x0021} \\
        L & Google Pixel 3a & 54 & 4 & b & NaN \\
        M & Apple iPhone XS Max & 2, 19 & 4 & b & \texttt{0x002d}, \texttt{0x402d} \\
        N & Apple iPhone 6 & 27 & 4 & b & \texttt{0x4021} \\
        O & OnePlus Nord & 62 & 4 & b & \texttt{0x01ad} \\
        Q & Huawei P10 & 98 & 4 & b & \texttt{0x01ad} \\
        R & Honor 9 & 287 & 4 & b & \texttt{0x0021} \\
        S & Xiaomi Redmi Note 7 & 64 & 4 & b & \texttt{0x012d} \\
        T & Xiaomi Redmi Note 9S & 17 & 4 & b & \texttt{0x01ad} \\
        U & Apple iPhone XR & 5 & 4 & b & \texttt{0x002d} \\
        V & Google Pixel 3a & 27 & 4 & b & NaN \\
        W & Apple iPhone 12 & 539 & 4 & b & \texttt{0x402d} \\
        X & Apple iPhone 7 & 7 & 4 & b & \texttt{0x402d} \\
        \bottomrule
        \multicolumn{6}{@{}p{\linewidth}@{}}{
            \scriptsize \textit{Note:} Some devices exhibited two different \ac{HT} capabilities information fields during the capture, resulting in two distinct Probe Request counts.
        }
    \end{tabularx}
\end{table}

Next, we describe the different pieces of information considered for \ac{MAC} address de-randomization.

\subsubsection{Static Characteristics (Decomposed \ac{HT} Capabilities)}
Among the \acp{IE} present in Probe Requests, the \ac{HT} capabilities field (\ac{IE} 45) consists of 28 bytes. Within this \ac{IE}, the \ac{HT} capabilities information accounts for 2 bytes, which is a 16-bit hexadecimal value (e.g., 0x01ad) that encodes 802.11 hardware features. In prior work, all 28 bytes were converted into decimal values, summed together, and treated as a single feature \cite{Pintor2024}. However, instead of doing this, we take only the \ac{HT} capabilities information, the A-MPDU parameters, and the \ac{HT} extended capabilities. One of the contributions of this work is to demonstrate that decomposing the \ac{HT} capabilities information adds granularity to the feature space, favoring de-randomization through clustering. Fig.~\ref{fig:ht_caps_decomposition} presents the bitwise representation of the \ac{HT} capabilities information.


\begin{figure}[!t]
\centering

\resizebox{\linewidth}{!}{
\renewcommand{\arraystretch}{1.5} 

\begin{tabular}{|*{14}{c|}}
\hline

\textbf{0} & \textbf{1} & \textbf{2--3} & \textbf{4} & \textbf{5} & \textbf{6} & \textbf{7} & \textbf{8--9} & \textbf{10} & \textbf{11} & \textbf{12} & \textbf{13} & \textbf{14} & \textbf{15} \\
\hline

\rotatebox{90}{\hspace*{1ex}\ac{LDPC}\hspace*{1ex}} &
\rotatebox{90}{\hspace*{1ex}Channel Width\hspace*{1ex}} &
\rotatebox{90}{\hspace*{1ex}SM Power Save\hspace*{1ex}} &
\rotatebox{90}{\hspace*{1ex}\ac{HT} Greenfield\hspace*{1ex}} &
\rotatebox{90}{\hspace*{1ex}Short GI 20\hspace*{1ex}} &
\rotatebox{90}{\hspace*{1ex}Short GI 40\hspace*{1ex}} &
\rotatebox{90}{\hspace*{1ex}Tx \ac{STBC}\hspace*{1ex}} &
\rotatebox{90}{\hspace*{1ex}Rx \ac{STBC}\hspace*{1ex}} &
\rotatebox{90}{\hspace*{1ex}\ac{HT}-del. BA\hspace*{1ex}} &
\rotatebox{90}{\hspace*{1ex}Max A-MSDU\hspace*{1ex}} &
\rotatebox{90}{\hspace*{1ex}DSSS/CCK 40\hspace*{1ex}} &
\rotatebox{90}{\hspace*{1ex}Reserved\hspace*{1ex}} &
\rotatebox{90}{\hspace*{1ex}40\,MHz Intolerant\hspace*{1ex}} &
\rotatebox{90}{\hspace*{1ex}L-SIG TXOP\hspace*{1ex}} \\

\hline
\end{tabular}
}
\caption{Bitwise structure of the 16-bit \ac{HT} capabilities information field. The first row indicates the bit positions and the second row, the subfields~\cite{ieee80211n}.}
\label{fig:ht_caps_decomposition}
\end{figure}

\subsubsection{Temporal Characteristics: IFAT}
Moving to timing features, it has been demonstrated that \ac{IFAT} characterizes the transmission behavior of a device \cite{Matte2016}.
In this work, the \ac{IFAT} is computed as the time difference between two consecutive Probe Requests from the same \ac{MAC} address:
\begin{equation}
\text{IFAT}_n = \text{Timestamp}_n - \text{Timestamp}_{n-1}.
\end{equation}

Frames are first grouped into bursts, which are consecutive frames that share the same source \ac{MAC} address and for which the elapsed time between frames is below 1 second. Although Probe Requests that belong to the same burst are transmitted only a few tens of milliseconds apart, the complete burst transmission lasts only a few hundred milliseconds. However, the active scanning process of a \ac{STA} may involve probing across multiple channels \cite{Praharenka2021}, causing the inter-burst time to span several seconds or, in some cases, minutes. Once the bursts are identified, the mean \ac{IFAT} is computed and assigned uniformly to all frames, including the first one, ensuring that every frame has an associated value.

\subsubsection{Spatial Characteristics (Simulated RSSI)}
As captured frames lack environmental context (they were extracted by filtering \ac{RSSI}), new \ac{RSSI} values are simulated to replace the original values using \ac{LNSM} with dynamic variance that depends on the distance from the STA to the sniffer. Using a fixed variance would be unrealistic because, in real-world scenarios, signal fluctuations depend on environmental conditions and vary with distance due to phenomena such as multipath, reflections, and shadowing. Therefore, we compute a signal decay based on a path loss index and use random Gaussian noise $N(0, \sigma(d)^2)$, where the dynamic variance $\sigma(d)$ is determined by applying a cubic function based on the device's distance ($\sigma(d) = ad^3 + bd^2 + cd + e$). The specific coefficients ($a = -0.0493$, $b = 0.3938$, $c = -0.5599$, $e = 0.4745$) are derived from \cite{Xu2010}'s indoor experiment (collected from a $15\times10$ hall). We assume that the signal captured from a given frame remains fixed during its entire transmission duration.

To simulate spatial characteristics, device positions are uniformly distributed within a circular area with a radius of 5 meters, and the Euclidean distance between the device and the sniffer is computed. Because a 5-meter radius simulates an area of 10-meter diameter, this scenario actually fits within the physical dimension of \cite{Xu2010}'s 15-by-10 meter experiment hall. However, as a limitation, these coefficients represent a specific environment and do not generalize to all spatial contexts. Assuming that devices remain static, three experimental configurations are evaluated: without \ac{RSSI} (Scenario 1), a single central sniffer (Scenario 2), and three sniffers arranged in a triangular topology (Scenario 3).

\subsubsection{Normalization}

Feature standardization is required before unsupervised clustering due to the varying scales and units (e.g., dBm, milliseconds, and binary flags) in the dataset. Without normalization, high-range variables like \ac{RSSI} would dominate the model, minimizing the contribution of smaller-range or binary variables like the \ac{HT} capabilities subfields.

\subsection{Clustering Algorithms} \label{sec:clustering_algorithms}
Three unsupervised clustering algorithms are evaluated: K-Means, DBSCAN, and OPTICS. On the one hand, K-Means has a lower computational complexity, $O(knT)$ (Algorithm~\ref{alg:kmeans_algorithm}), where $k$ is the number of clusters, $n$ is the number of samples, and $T$ is the number of iterations. However, it requires the number of clusters to be predefined.
This limitation is addressed by DBSCAN (Algorithm~\ref{alg:dbscan_algorithm}), which automatically identifies clusters. Moreover, it does not force all samples into clusters; samples with uncertain assignments can instead be treated as outliers. Moreover, DBSCAN assumes that clusters exhibit similar densities. Since devices may generate different numbers of Probe Frames, OPTICS is also considered (Algorithm~\ref{alg:optics_algorithm}), as it better handles clusters with varying densities. Both DBSCAN and OPTICS have a worst-case computational complexity of $O(n^2)$~\cite{scikit-learn}.

\begin{algorithm}[!t]
\caption{K-Means Algorithm}
\label{alg:kmeans_algorithm}
\small
\begin{algorithmic}[1]
\REQUIRE Dataset $X$, number of clusters $k$
\ENSURE Cluster assignments $L$ and centroids $C$
\STATE Randomly initialize the centroids $C = \{c_1, c_2, \dots, c_k\}$
\STATE Initialize an empty labels array $L = \{l_1, l_2, \dots, l_n\}$
\REPEAT
\FOR{each data point $x_i \in X$}
\STATE Calculate the distance of $x_i$ to all centroids $c_j \in C$
\STATE Assign $x_i$ to the closest centroid $c_j \in C$
\STATE Update label $l_i = j$
\ENDFOR
\FOR{each cluster $j \in \{1, \dots, k\}$}
\STATE Update centroids $c_j \leftarrow \frac{1}{|S_j|} \sum_{x_i \in S_j} x_i$
\ENDFOR
\UNTIL{centroids do not change or maximum iterations reached}
\RETURN $L, C$
\end{algorithmic}
\end{algorithm}

\begin{algorithm}[!t]
\caption{DBSCAN Algorithm}
\label{alg:dbscan_algorithm}
\small
\begin{algorithmic}[1]
\REQUIRE Dataset $X$, radius $\epsilon$, minimum points $s_\text{min}$
\ENSURE Cluster assignment $L$
\STATE $cluster\_id \leftarrow 0$
\STATE Compute $\epsilon$-neighborhood $N[p]$ for all $p \in X$
\STATE $core\_point \leftarrow \{p \in X \mid |N[p]| \geq s_\text{min}\}$
\STATE Initialize $L[p] \leftarrow -1$ for all $p \in X$
\FOR{each $p \in core\_point$ \textbf{where} $L[p] = -1$}
    \STATE $L[p] \leftarrow cluster\_id$
    \STATE Initialize stack $S \leftarrow [p]$
    \WHILE{$S$ is not empty}
        \STATE $q \leftarrow S.\text{pop}()$
        \IF{$q \in core\_point$}
            \FOR{each $n \in N[q]$ \textbf{where} $L[n] = -1$}
                \STATE $L[n] \leftarrow cluster\_id$
                \STATE $S.\text{push}(n)$
            \ENDFOR
        \ENDIF
    \ENDWHILE
    \STATE $cluster\_id \leftarrow cluster\_id + 1$
\ENDFOR
\RETURN $L$
\end{algorithmic}
\end{algorithm}

\begin{algorithm}[!t]
\caption{OPTICS Algorithm}
\label{alg:optics_algorithm}
\small
\begin{algorithmic}[1]
\REQUIRE Dataset $X$, maximum radius $\epsilon_\text{max}$, minimum points $s_\text{min}$,
\ENSURE Cluster assignments $L$

\STATE Compute $core\_dist[p]$ for all $p \in X$
\STATE \text{Initialize arrays:} $reach \leftarrow \infty$, $processed \leftarrow 0$, $L \leftarrow -1$
\STATE \text{Initialize list} $Order \leftarrow []$
\STATE $cluster\_id \leftarrow 0$

\WHILE{there exists $p \in X$ \textbf{where} $processed[p] = 0$}
    \STATE $p \leftarrow$ unprocessed point with the minimum $reach$ value
    \STATE $processed[p] \leftarrow 1$
    \STATE Append $p$ to $Order$
    
    \IF{$core\_dist[p] \leq \epsilon_\text{max}$}
        \FOR{each unprocessed neighbor $q$ within $\epsilon_\text{max}$}
            \STATE $new\_reach \leftarrow \max(core\_dist[p], \text{Distance}(p, q))$
            \STATE $reach[q] \leftarrow \min(reach[q], new\_reach)$
        \ENDFOR
    \ENDIF
\ENDWHILE

\FOR{each $p \in Order$}
    \IF{$reach[p] > \epsilon$ \textbf{and} $core\_dist[p] \leq \epsilon$}
        \STATE $cluster\_id \leftarrow cluster\_id + 1$ \hfill \textit{// New cluster}
        \STATE $L[p] \leftarrow cluster\_id$
    \ELSIF{$reach[p] \leq \epsilon$}
        \STATE $L[p] \leftarrow cluster\_id$ \hfill \textit{// Current cluster}
    \ENDIF
\ENDFOR

\RETURN $L$
\end{algorithmic}
\end{algorithm}

\subsection{Evaluation Metrics} \label{sec:evaluation_metrics}
Unsupervised clustering performance is assessed using four metrics: Global Accuracy, Precision, Recall, and Individual Device Accuracy. Although the clustering algorithms used are unsupervised, we evaluate their performance using metric associated with supervised learning. This was possibly due to the availability of ground-truth labels in the dataset, which allows us to directly compare the unsupervised cluster assignments against the true device identities.

The metrics are defined as follows:
\begin{itemize}
    \item \textbf{Global Accuracy:} The ratio of correctly classified Probe Requests to the total number of Probe Requests in the dataset.
    \item \textbf{Individual Accuracy:} Follows the same principle as global accuracy, but for each specific device.
    \item \textbf{Precision:} The ratio of Probe Requests correctly assigned to a device to the total number of Probe Requests assigned to that device's mapped cluster. The final metric is the unweighted mean of the precision scores across all devices. This ensures that every device contributes equally to the final score, regardless of its number of Probe Requests.
    \begin{equation}
      \text{Precision} = \frac{1}{M} \sum_{i=1}^{M} \frac{\rm{TP}_i}{\rm{TP}_i + \rm{FP}_i},  
    \end{equation}
    where $\rm{TP}$ stands for True Positives, $\rm{FP}$ for False Positives, and $M$ for the total number of devices.
    \item \textbf{Recall:} The ratio of correctly clustered Probe Requests to the total number of Probe Requests actually transmitted by that device. Just as with precision, the final recall is the unweighted mean of the recall scores across all devices.
    \begin{equation}
        \text{Recall} = \frac{1}{M} \sum_{i=1}^{M} \frac{\rm{TP}_i}{\rm{TP}_i + \rm{FN}_i},
    \end{equation} 
    where $\rm{FN}$ stands for False Negatives.  
\end{itemize}{}

Because the clusters generated by unsupervised algorithms are not labeled, it is necessary to map the predicted clusters to the corresponding ground-truth labels in the dataset. This is achieved using the Hungarian algorithm, which efficiently finds the assignment that maximizes the number of correct matches without requiring the evaluation of all possible permutations \cite{Gil-Aluja1998}. It should be noted that outlier samples labeled as \texttt{-1} by DBSCAN and OPTICS are excluded from the contingency matrix used for the cluster-to-label mapping. Because \texttt{-1} samples are not part of any cluster, including these instances disrupts the algorithm's ability to find the optimal mapping. However, these outliers are included in the computation of all evaluation metrics, since any Probe Requests labeled as \texttt{-1} represent a mistake of the algorithm to assign them to their correct device cluster.

\section{Experimental Setup} \label{sec:experimental_setup}
This section outlines the experimental configurations used to evaluate the clustering performance. Before doing so, we first justify the features selected using a feature importance analysis.

\subsection{Feature Importance Analysis} \label{sec:feature_importance}

Feature importance is analyzed using an \ac{RF}, which evaluates the contribution of temporal, spatial, and protocol-level features. Specifically, it focuses on three key comparisons to justify the decomposition of \ac{HT} capabilities information, the impact of temporal information (\ac{IFAT}), and the influence of spatial information (\ac{RSSI}). While not shown here for the sake of space, the results consistently showed the great importance of specific \ac{HT} capabilities subfields. Additionally, \ac{IFAT} was revealed as one of the top three most important features. Finally, when \ac{RSSI} values were incorporated, whether using one or three signals, they always reached the highest position in the feature hierarchy.

Based on the feature importance results obtained from the \ac{RF} models across miscellaneous scenarios, a final set of features was selected for the unsupervised clustering stage. The ones that consistently showed zero importance, such as \texttt{channel}, \texttt{rates}, \texttt{ext\_rates}, and \texttt{vht\_caps}, were excluded from the final feature set. Nevertheless, special consideration was given to the \texttt{ssid} field. Although it often returned a near-zero importance score (because most Probe Requests in the dataset used wildcard \acp{SSID}), it was retained. This decision was driven by the fact that some devices still broadcast specific \acp{SSID}, which can serve as an additional feature that could improve the identification of devices that share similar \ac{HT} specifications. Additionally, the derived feature \texttt{oui\_from\_mac} is obtained from the first 24 bits of the resolved \ac{MAC} address to serve as an extra vendor identifier alongside \texttt{oui}.
The final consolidated feature set used for clustering includes:
\begin{itemize}
    \item Temporal markers: \texttt{time} and \texttt{time\_diff} (\ac{IFAT}).
    \item Hardware identifiers: Decomposed \ac{HT} subfields (or raw \texttt{ht\_caps}), \texttt{htex\_caps}, and \texttt{ampdu}.
    \item Vendor information: \texttt{oui} and \texttt{oui\_from\_mac}.
    \item Spatial information: Simulated \ac{RSSI} (single/tri-sniffer).
\end{itemize}

\subsection{Scenarios and Configurations} 
Using the feature set justified above, we construct 12 distinct experimental configurations. Specifically, three spatial scenarios with different feature sets were considered to reflect different levels of an eavesdropper's infrastructure and physical tracking capabilities:

\begin{itemize}
    \item Scenario 1 (Without \ac{RSSI}): Passive attack relying solely on Frame headers and \ac{IFAT} timing.
    \item Scenario 2 (Single-Sniffer, \ac{SRSSI}): One monitoring node providing a single \ac{RSSI} per Frame.
    \item Scenario 3 (Multi-Sniffer, \ac{MSRSSI}): Three nodes in a triangular topology providing three \ac{RSSI} values per Frame.
\end{itemize}

Within each scenario, four distinct setups are tested to evaluate the contribution of specific features:

\begin{itemize}
    \item Raw/Decomposed Feature Representation: Comparing the use of the raw Hexadecimal \texttt{ht\_caps} string against the decomposed binary subfields.
    \item With/Without Temporal Information: Evaluating the models both with and without the \ac{IFAT}.
\end{itemize}

So in total, there are 12 experimental configurations illustrated in Table~\ref{tab:scenarios}. Each configuration is evaluated across four device densities (5, 10, 15, and 22 devices). For the 5, 10, and 15 devices, five random subsets of devices are generated without replacement, and results are averaged across these iterations. For the 22-device case, the full dataset is used, so only a single result is obtained, and no averaging is required.

\begin{table}[t]
\centering
\small
\renewcommand{\arraystretch}{1.1}
\caption{Experimental configurations: spatial scenarios and feature setups.}
\label{tab:scenarios}

\resizebox{\columnwidth}{!}{%
\begin{tabular}{c l c l}
    \toprule
    \multicolumn{2}{c}{\textbf{Spatial Scenario}}
    & \multicolumn{2}{c}{\textbf{Feature Setup}} \\
    \cmidrule(r){1-2}\cmidrule(l){3-4}
    \textbf{Scenario} & \textbf{Description} & \textbf{ID} & \textbf{Description} \\
    \midrule

    \multirow{4}{*}{1} & \multirow{4}{*}{Without \ac{RSSI}}
        & S1.1 & Raw \texttt{ht\_caps}, w/o \ac{IFAT} \\
    & & S1.2 & Raw \texttt{ht\_caps}, w/ \ac{IFAT} \\
    & & S1.3 & Decomposed subfields, w/o \ac{IFAT} \\
    & & S1.4 & Decomposed subfields, w/ \ac{IFAT} \\

    \midrule

    \multirow{4}{*}{2} & \multirow{4}{*}{Single-sniffer}
        & S2.1 & Raw \texttt{ht\_caps}, w/o \ac{IFAT} \\
    & & S2.2 & Raw \texttt{ht\_caps}, w/ \ac{IFAT} \\
    & & S2.3 & Decomposed subfields, w/o \ac{IFAT} \\
    & & S2.4 & Decomposed subfields, w/ \ac{IFAT} \\

    \midrule

    \multirow{4}{*}{3} & \multirow{4}{*}{Multi-sniffer}
        & S3.1 & Raw \texttt{ht\_caps}, w/o \ac{IFAT} \\
    & & S3.2 & Raw \texttt{ht\_caps}, w/ \ac{IFAT} \\
    & & S3.3 & Decomposed subfields, w/o \ac{IFAT} \\
    & & S3.4 & Decomposed subfields, w/ \ac{IFAT} \\
    \bottomrule
\end{tabular}%
}
\end{table}

\section{Results and Discussion} \label{sec:results}

This section presents the evaluation of the proposed unsupervised clustering algorithms (K-Means, DBSCAN, and OPTICS) across three spatial configurations and different device densities. Next, we analyze the individual accuracy per device, and finally, we summarize the main challenges identified during the evaluation and potential directions for future work.

\subsection{Clustering Performance}
Table \ref{tab:hyperparameters} summarizes the specific parameter configurations used to evaluate these algorithms.

\begin{table}[!t]
    \centering
    \caption{Hyperparameter configurations for the unsupervised clustering algorithms.}
    \label{tab:hyperparameters}
    \footnotesize 
    \renewcommand{\arraystretch}{1.2}

    \begin{tabularx}{\linewidth}{@{} l l X @{}}
        \toprule
        \textbf{Algorithm} & \textbf{Hyperparameter} & \textbf{Configured Value} \\
        \midrule
        \textbf{K-Means} & Clusters ($k$) & 5, 10, 15, or 22 (Dataset dependent) \\
        \textbf{DBSCAN} & $min\_samples$ & 5 \\
        \textbf{DBSCAN} & Epsilon ($\epsilon$) & Dynamic (Optimised) \\
        \textbf{OPTICS} & $min\_samples$ & 5 \\
        \textbf{OPTICS} & Steepness ($\xi$) & 0.2 \\
        \bottomrule
    \end{tabularx}
\end{table}

\subsubsection{Scenario 1 (Without \ac{RSSI})}
\begin{figure*}[t]
    \centering
    \begin{tikzpicture}
        \begin{axis}[
            title={\textbf{Raw HT capabilities information}},
            xlabel={Number of Devices},
            ylabel={Accuracy (\%)},
            xmin=3, xmax=25,
            ymin=0, ymax=100,
            xtick={5,10,15,22},
            ytick distance=10,
            ymajorgrids=true,
            grid style=dashed,
            width=0.44\textwidth,
            height=5.7cm,
            every node near coord/.append style={font=\tiny,
                /pgf/number format/precision=0},
        ]
        \addplot[color=blue, mark=square*, thick,
            nodes near coords,
            every node near coord/.append style={xshift=-2pt,anchor=east}]
            coordinates {(5,62.028)(10,52.530)(15,46.060)(22,44.800)};
        \addplot[color=red, mark=*, thick,
            nodes near coords,
            every node near coord/.append style={yshift=2pt,anchor=south}]
            coordinates {(5,79.524)(10,80.386)(15,84.000)(22,82.280)};
        \addplot[color=green!60!black, mark=triangle*, thick,
            nodes near coords,
            every node near coord/.append style={xshift=2pt,anchor=west}]
            coordinates {(5,12.858)(10,25.500)(15,53.760)(22,57.520)};
        \addplot[color=blue, mark=square, dashed, thick,
            nodes near coords,
            every node near coord/.append style={yshift=2pt,xshift=-2pt,
                anchor=south east}]
            coordinates {(5,71.714)(10,57.614)(15,48.558)(22,46.440)};
        \addplot[color=red, mark=o, dashed, thick,
            nodes near coords,
            every node near coord/.append style={yshift=-2pt,anchor=north}]
            coordinates {(5,78.832)(10,79.640)(15,81.030)(22,79.670)};
        \addplot[color=green!60!black, mark=triangle, dashed, thick,
            nodes near coords,
            every node near coord/.append style={yshift=-2pt,xshift=2pt,
                anchor=north west}]
            coordinates {(5,4.050)(10,22.760)(15,39.920)(22,55.040)};
        \end{axis}
    \end{tikzpicture}
    \hfill
    \begin{tikzpicture}
        \begin{axis}[
            title={\textbf{Decomposed HT capabilities information}},
            xlabel={Number of Devices},
            xmin=3, xmax=25,
            ymin=0, ymax=100,
            xtick={5,10,15,22},
            ytick distance=10,
            legend style={at={(1.05,0.5)}, anchor=west,
                nodes={scale=0.62,transform shape}},
            ymajorgrids=true,
            grid style=dashed,
            width=0.44\textwidth,
            height=5.7cm,
            every node near coord/.append style={font=\tiny,
                /pgf/number format/precision=0},
        ]
        \addplot[color=blue, mark=square*, thick,
            nodes near coords,
            every node near coord/.append style={xshift=-2pt,anchor=east}]
            coordinates {(5,70.052)(10,62.950)(15,50.478)(22,51.270)};
        \addlegendentry{K-Means}
        \addplot[color=red, mark=*, thick,
            nodes near coords,
            every node near coord/.append style={yshift=2pt,anchor=south}]
            coordinates {(5,86.536)(10,89.520)(15,89.738)(22,87.320)};
        \addlegendentry{DBSCAN}
        \addplot[color=green!60!black, mark=triangle*, thick,
            nodes near coords,
            every node near coord/.append style={xshift=2pt,anchor=west}]
            coordinates {(5,21.672)(10,35.636)(15,74.986)(22,64.340)};
        \addlegendentry{OPTICS}
        \addplot[color=blue, mark=square, dashed, thick,
            nodes near coords,
            every node near coord/.append style={yshift=2pt,xshift=-2pt,
                anchor=south east}]
            coordinates {(5,78.462)(10,67.690)(15,52.678)(22,50.880)};
        \addlegendentry{K-Means (\ac{IFAT})}
        \addplot[color=red, mark=o, dashed, thick,
            nodes near coords,
            every node near coord/.append style={yshift=-2pt,anchor=north}]
            coordinates {(5,86.106)(10,83.754)(15,87.314)(22,86.480)};
        \addlegendentry{DBSCAN (\ac{IFAT})}
        \addplot[color=green!60!black, mark=triangle, dashed, thick,
            nodes near coords,
            every node near coord/.append style={yshift=-2pt,xshift=2pt,
                anchor=north west}]
            coordinates {(5,5.342)(10,27.818)(15,37.720)(22,62.140)};
        \addlegendentry{OPTICS (\ac{IFAT})}
        \end{axis}
    \end{tikzpicture}
    \caption{Clustering accuracy vs.\ device count for Scenario~1
    (without \ac{RSSI}). Left: raw hexadecimal \ac{HT} capabilities information (S1.1, S1.2).
    Right: bitwise-decomposed \ac{HT} capabilities information (S1.3, S1.4).
    Solid lines: without \ac{IFAT}; dashed lines: with \ac{IFAT}.}
    \label{fig:scenario1_performance}
\end{figure*}
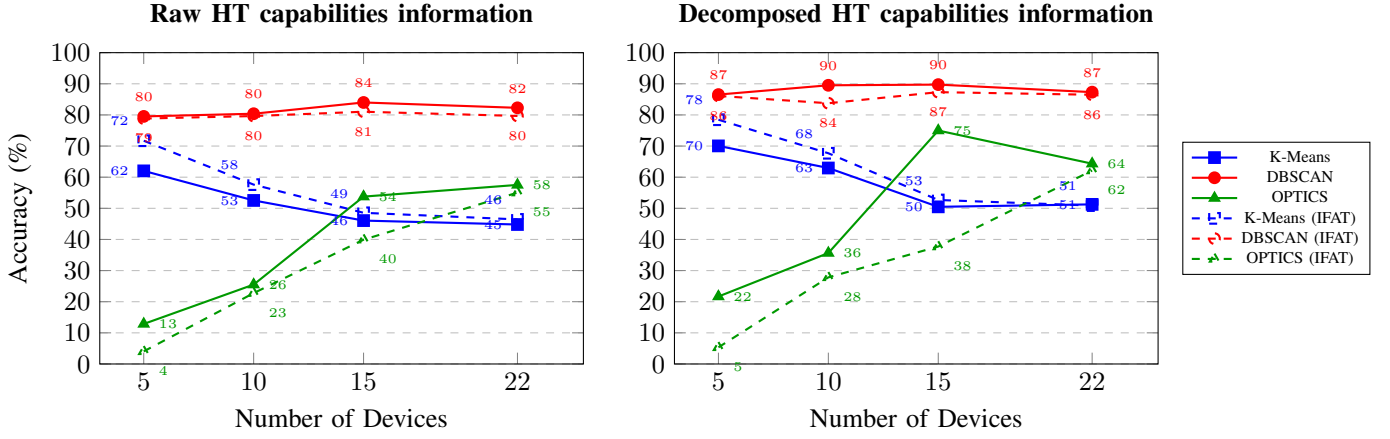

Scenario 1 evaluates the baseline clustering performance, where the assumption is that only the Probe Frame header and temporal information are available. Because they must rely entirely on the fingerprint of the device, this scenario is susceptible to possible hardware collisions when devices from the same manufacturer are present in the same environment. The overall results of this scenario across different device densities are illustrated in Fig. \ref{fig:scenario1_performance}. It was observed that decomposing \ac{HT} capabilities information into individual subfields consistently improved clustering performance compared to using the raw hexadecimal representation. The impact of this decomposition was most noticeable in K-Means, where average accuracy increased significantly, though its performance still progressively decreased as the number of devices increased and the feature space became more crowded. DBSCAN demonstrated greater robustness under these decomposed setups, yielding an average global accuracy of 88.28\%. In contrast, OPTICS generally underperformed in comparison to both K-Means and DBSCAN. Regarding temporal data, \ac{IFAT} often improved cluster separation for K-Means. Conversely, \ac{IFAT} negatively affected DBSCAN's performance, as the algorithm interprets this temporal variability as noise, which weakens density consistency and can fragment otherwise stable clusters.

\subsubsection{Scenario 2 (\ac{SRSSI})}
\begin{figure*}[t]
    \centering
    \begin{tikzpicture}
        \begin{axis}[
            title={\textbf{Raw HT capabilities information}},
            xlabel={Number of Devices},
            ylabel={Accuracy (\%)},
            xmin=3, xmax=25,
            ymin=0, ymax=100,
            xtick={5,10,15,22},
            ytick distance=10,
            ymajorgrids=true,
            grid style=dashed,
            width=0.44\textwidth,
            height=5.7cm,
            every node near coord/.append style={font=\tiny,
                /pgf/number format/precision=0},
        ]
        \addplot[color=blue, mark=square*, thick,
            nodes near coords,
            every node near coord/.append style={xshift=-2pt,anchor=east}]
            coordinates {(5,67.974)(10,50.768)(15,47.542)(22,42.980)};
        \addplot[color=red, mark=*, thick,
            nodes near coords,
            every node near coord/.append style={yshift=2pt,anchor=south}]
            coordinates {(5,93.084)(10,92.672)(15,86.250)(22,84.200)};
        \addplot[color=green!60!black, mark=triangle*, thick,
            nodes near coords,
            every node near coord/.append style={xshift=2pt,anchor=west}]
            coordinates {(5,30.820)(10,45.924)(15,47.446)(22,63.730)};
        \addplot[color=blue, mark=square, dashed, thick,
            nodes near coords,
            every node near coord/.append style={yshift=2pt,xshift=-2pt,
                anchor=south east}]
            coordinates {(5,63.514)(10,53.808)(15,51.168)(22,45.540)};
        \addplot[color=red, mark=o, dashed, thick,
            nodes near coords,
            every node near coord/.append style={yshift=-2pt,anchor=north}]
            coordinates {(5,86.786)(10,88.872)(15,84.338)(22,83.170)};
        \addplot[color=green!60!black, mark=triangle, dashed, thick,
            nodes near coords,
            every node near coord/.append style={yshift=-2pt,xshift=2pt,
                anchor=north west}]
            coordinates {(5,30.364)(10,38.138)(15,47.548)(22,61.840)};
        \end{axis}
    \end{tikzpicture}
    \hfill
    \begin{tikzpicture}
        \begin{axis}[
            title={\textbf{Decomposed HT capabilities information}},
            xlabel={Number of Devices},
            xmin=3, xmax=25,
            ymin=0, ymax=100,
            xtick={5,10,15,22},
            ytick distance=10,
            legend style={at={(1.05,0.5)}, anchor=west,
                nodes={scale=0.62,transform shape}},
            ymajorgrids=true,
            grid style=dashed,
            width=0.44\textwidth,
            height=5.7cm,
            every node near coord/.append style={font=\tiny,
                /pgf/number format/precision=0},
        ]
        \addplot[color=blue, mark=square*, thick,
            nodes near coords,
            every node near coord/.append style={xshift=-2pt,anchor=east}]
            coordinates {(5,86.930)(10,62.598)(15,56.244)(22,53.910)};
        \addlegendentry{K-Means}
        \addplot[color=red, mark=*, thick,
            nodes near coords,
            every node near coord/.append style={yshift=2pt,anchor=south}]
            coordinates {(5,94.620)(10,92.164)(15,90.454)(22,88.600)};
        \addlegendentry{DBSCAN}
        \addplot[color=green!60!black, mark=triangle*, thick,
            nodes near coords,
            every node near coord/.append style={xshift=2pt,anchor=west}]
            coordinates {(5,29.384)(10,46.444)(15,50.278)(22,63.730)};
        \addlegendentry{OPTICS}
        \addplot[color=blue, mark=square, dashed, thick,
            nodes near coords,
            every node near coord/.append style={yshift=2pt,xshift=-2pt,
                anchor=south east}]
            coordinates {(5,92.132)(10,64.442)(15,55.524)(22,53.630)};
        \addlegendentry{K-Means (\ac{IFAT})}
        \addplot[color=red, mark=o, dashed, thick,
            nodes near coords,
            every node near coord/.append style={yshift=-2pt,anchor=north}]
            coordinates {(5,89.892)(10,91.606)(15,88.420)(22,87.340)};
        \addlegendentry{DBSCAN (\ac{IFAT})}
        \addplot[color=green!60!black, mark=triangle, dashed, thick,
            nodes near coords,
            every node near coord/.append style={yshift=-2pt,xshift=2pt,
                anchor=north west}]
            coordinates {(5,27.768)(10,39.152)(15,53.284)(22,61.750)};
        \addlegendentry{OPTICS (\ac{IFAT})}
        \end{axis}
    \end{tikzpicture}
    \caption{Clustering accuracy vs.\ device count for Scenario~2
    (single-sniffer \ac{RSSI}). Left: raw \ac{HT} capabilities information (S2.1, S2.2).
    Right: decomposed \ac{HT} capabilities information (S2.3, S2.4).
    Solid: no \ac{IFAT}; dashed: with \ac{IFAT}.}
    \label{fig:scenario2_performance}
\end{figure*}
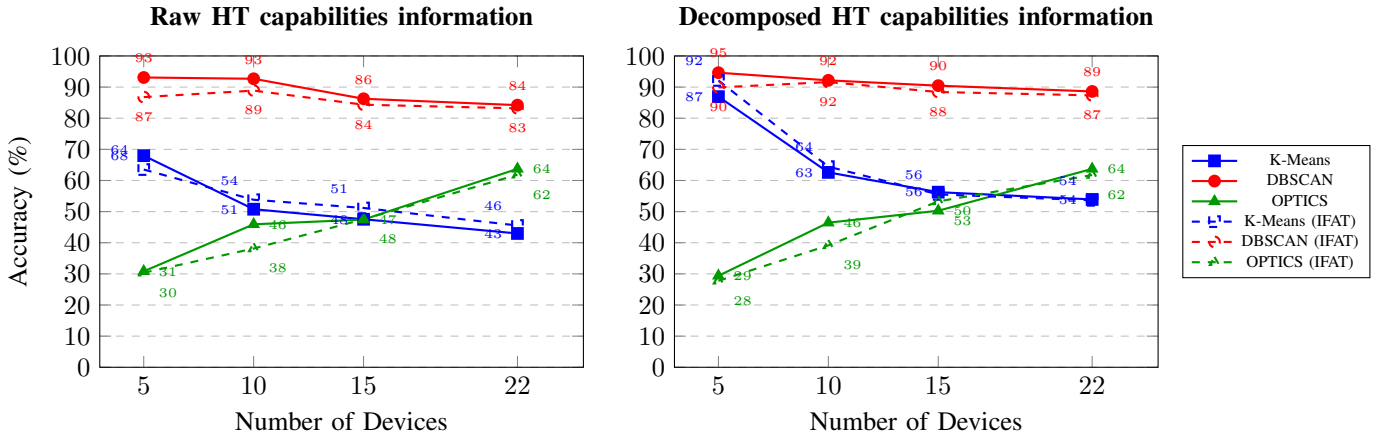

Scenario 2 introduces spatial information into the clustering environment by simulating a single \ac{RSSI} value. As a spatial feature, \ac{RSSI} cannot give the exact device's position, but it can provide an estimated distance, which could be advantageous to the clustering models. The clustering performance under this second scenario is illustrated in Fig. \ref{fig:scenario2_performance}. The addition of this spatial data improved the performance of all three clustering algorithms compared to scenario 1. The necessity of decomposing the \ac{HT} capabilities information remains evident even with the inclusion of spatial data. With the decomposed setup, DBSCAN demonstrated strong robustness and achieved the highest overall performance, reaching an accuracy of 88.60\% for 22 devices. K-Means also benefited from the inclusion of \ac{SRSSI} in less crowded environments, but it failed to maintain this improvement as the number of devices increased. On the other hand, OPTICS showed only slight improvements, exhibiting a performance significantly inferior to the other two algorithms. For DBSCAN, the inclusion of \ac{IFAT} remained unfavorable, as it introduced noise despite the better separation provided by \ac{HT} capabilities information decomposition and \ac{SRSSI}.

\subsubsection{Scenario 3 (\ac{MSRSSI})}
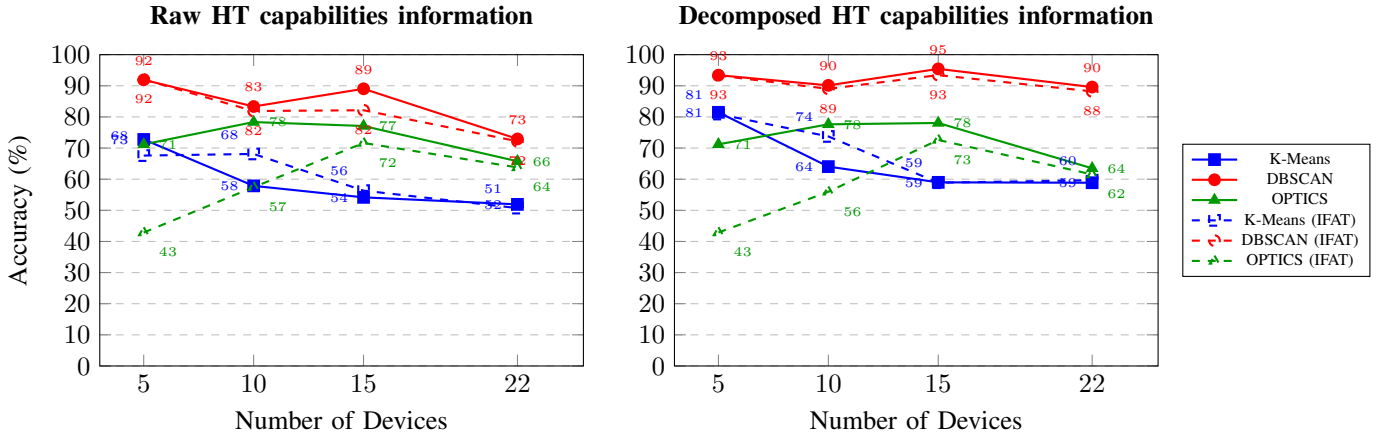
\begin{figure*}[t]
    \centering
    \begin{tikzpicture}
        \begin{axis}[
            title={\textbf{Raw HT capabilities information}},
            xlabel={Number of Devices},
            ylabel={Accuracy (\%)},
            xmin=3, xmax=25,
            ymin=0, ymax=100,
            xtick={5,10,15,22},
            ytick distance=10,
            ymajorgrids=true,
            grid style=dashed,
            width=0.44\textwidth,
            height=5.7cm,
            every node near coord/.append style={font=\tiny,
                /pgf/number format/precision=0},
        ]
        \addplot[color=blue, mark=square*, thick,
            nodes near coords,
            every node near coord/.append style={xshift=-2pt,anchor=east}]
            coordinates {(5,72.760)(10,57.856)(15,54.164)(22,51.910)};
        \addplot[color=red, mark=*, thick,
            nodes near coords,
            every node near coord/.append style={yshift=2pt,anchor=south}]
            coordinates {(5,91.914)(10,83.350)(15,89.036)(22,72.930)};
        \addplot[color=green!60!black, mark=triangle*, thick,
            nodes near coords,
            every node near coord/.append style={xshift=2pt,anchor=west}]
            coordinates {(5,71.192)(10,78.352)(15,77.084)(22,65.760)};
        \addplot[color=blue, mark=square, dashed, thick,
            nodes near coords,
            every node near coord/.append style={yshift=2pt,xshift=-2pt,
                anchor=south east}]
            coordinates {(5,67.584)(10,68.108)(15,56.328)(22,50.710)};
        \addplot[color=red, mark=o, dashed, thick,
            nodes near coords,
            every node near coord/.append style={yshift=-2pt,anchor=north}]
            coordinates {(5,92.008)(10,81.900)(15,82.146)(22,72.180)};
        \addplot[color=green!60!black, mark=triangle, dashed, thick,
            nodes near coords,
            every node near coord/.append style={yshift=-2pt,xshift=2pt,
                anchor=north west}]
            coordinates {(5,42.866)(10,57.452)(15,71.626)(22,63.780)};
        \end{axis}
    \end{tikzpicture}
    \hfill
    \begin{tikzpicture}
        \begin{axis}[
            title={\textbf{Decomposed HT capabilities information}},
            xlabel={Number of Devices},
            xmin=3, xmax=25,
            ymin=0, ymax=100,
            xtick={5,10,15,22},
            ytick distance=10,
            legend style={at={(1.05,0.5)}, anchor=west,
                nodes={scale=0.62,transform shape}},
            ymajorgrids=true,
            grid style=dashed,
            width=0.44\textwidth,
            height=5.7cm,
            every node near coord/.append style={font=\tiny,
                /pgf/number format/precision=0},
        ]
        \addplot[color=blue, mark=square*, thick,
            nodes near coords,
            every node near coord/.append style={xshift=-2pt,anchor=east}]
            coordinates {(5,81.480)(10,64.042)(15,58.986)(22,58.860)};
        \addlegendentry{K-Means}
        \addplot[color=red, mark=*, thick,
            nodes near coords,
            every node near coord/.append style={yshift=2pt,anchor=south}]
            coordinates {(5,93.414)(10,90.138)(15,95.432)(22,89.600)};
        \addlegendentry{DBSCAN}
        \addplot[color=green!60!black, mark=triangle*, thick,
            nodes near coords,
            every node near coord/.append style={xshift=2pt,anchor=west}]
            coordinates {(5,71.192)(10,77.624)(15,78.058)(22,63.500)};
        \addlegendentry{OPTICS}
        \addplot[color=blue, mark=square, dashed, thick,
            nodes near coords,
            every node near coord/.append style={yshift=2pt,xshift=-2pt,
                anchor=south east}]
            coordinates {(5,80.924)(10,73.734)(15,58.866)(22,59.670)};
        \addlegendentry{K-Means (\ac{IFAT})}
        \addplot[color=red, mark=o, dashed, thick,
            nodes near coords,
            every node near coord/.append style={yshift=-2pt,anchor=north}]
            coordinates {(5,93.312)(10,89.016)(15,93.428)(22,88.210)};
        \addlegendentry{DBSCAN (\ac{IFAT})}
        \addplot[color=green!60!black, mark=triangle, dashed, thick,
            nodes near coords,
            every node near coord/.append style={yshift=-2pt,xshift=2pt,
                anchor=north west}]
            coordinates {(5,42.866)(10,55.958)(15,72.604)(22,61.530)};
        \addlegendentry{OPTICS (\ac{IFAT})}
        \end{axis}
    \end{tikzpicture}
    \caption{Clustering accuracy vs.\ device count for Scenario~3
    (multi-sniffer \ac{RSSI}). Left: raw \ac{HT} capabilities information (S3.1, S3.2).
    Right: decomposed \ac{HT} capabilities information (S3.3, S3.4).
    Solid: no \ac{IFAT}; dashed: with \ac{IFAT}.}
    \label{fig:scenario3_performance}
\end{figure*}

Scenario 3 introduces a more robust spatial configuration into the clustering environment by simulating three \ac{RSSI} values from three different sniffers. Adding two additional \ac{RSSI} values provides more spatial information to better determine the physical location of the Probe Frame in the space. The effect of incorporating \ac{RSSI} measurements from three spatially separated sniffers on clustering accuracy is detailed in Fig. \ref{fig:scenario3_performance}. Consequently, the inclusion of multi-sniffer \ac{RSSI} measurements achieved the best performance in the three scenarios. The use of three \ac{RSSI} signals allows the algorithms to more effectively cluster Probe Frames originating from the same physical space independently of the randomized \ac{MAC} address or devices with identical \ac{HT} capabilities information. Even with the inclusion of three \ac{RSSI} signals, decomposing the \ac{HT} capabilities information from a raw 16-bit hexadecimal string to individual subfields remains critical to obtain high accuracy. In the 22-device environment, DBSCAN reported an accuracy of 89.60\% with decomposed \ac{HT} capabilities information, which corresponds to a 22.85\% increase compared to not using decomposition. K-Means achieved better results than in scenarios 1 and 2, but it still failed to correctly cluster approximately 40\% of the Probe Frames in the 22-device environment. OPTICS also showed its best performance in this scenario, but its overall accuracy remained significantly inferior to DBSCAN.

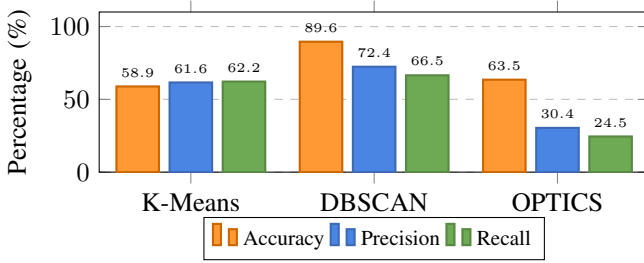
\begin{figure}[t]
    \centering
    \begin{tikzpicture}
        \begin{axis}[
            ybar=4pt,
            bar width=16pt,
            width=\columnwidth,
            height=3.7cm,
            enlarge x limits=0.25,
            legend style={at={(0.5,-0.28)}, anchor=north,
                legend columns=-1, font=\footnotesize},
            ylabel={Percentage (\%)},
            symbolic x coords={K-Means, DBSCAN, OPTICS},
            xtick=data,
            nodes near coords={%
                \pgfmathprintnumber[fixed,precision=1]{\pgfplotspointmeta}},
            nodes near coords align={vertical},
            every node near coord/.append style={font=\tiny},
            ymin=0, ymax=110,
            ymajorgrids=true,
            grid style=dashed,
        ]
        \definecolor{barBlue}{HTML}{4B85E3}
        \definecolor{barGreen}{HTML}{6EAA54}
        \addplot[fill=orange!80, draw=orange!80!black, thick]
            coordinates {(K-Means,58.86)(DBSCAN,89.60)(OPTICS,63.50)};
        \addplot[fill=barBlue, draw=barBlue!80!black, thick]
            coordinates {(K-Means,61.6)(DBSCAN,72.4)(OPTICS,30.4)};
        \addplot[fill=barGreen, draw=barGreen!80!black, thick]
            coordinates {(K-Means,62.2)(DBSCAN,66.5)(OPTICS,24.5)};
        \legend{Accuracy, Precision, Recall}
        \end{axis}
    \end{tikzpicture}
    \caption{Global accuracy, precision, and recall for all three algorithms under setup S3.3 (22 devices, decomposed \ac{HT} capabilities information, three \ac{RSSI} signals, no \ac{IFAT}).}
    \label{fig:metrics_s3}
\end{figure}

Fig.~\ref{fig:metrics_s3} presents global accuracy, precision, and recall for the 22-device environment under S3.3. DBSCAN achieves 72.4\% precision and 66.5\% recall, substantially ahead of K-Means (61.6\% and 62.2\%) and OPTICS (30.4\% and 24.5\%).

\subsection{Individual Device Analysis}
\label{sec:individual_device_analysis}

\begin{table}[t]
\caption{Per-device accuracy under S3.3 (22 devices, DBSCAN).
Shading: \colorbox{green!35}{$\geq$80\%}, \colorbox{yellow!40}{50--79\%}, \colorbox{orange!40}{1--49\%}, \colorbox{red!30}{0\%}. The third column for each device indicates the number of matched/total Probe Requests.}
\label{tab:individual_accuracies}
\centering
\small
\renewcommand{\arraystretch}{1.08}
\setlength{\tabcolsep}{4pt} 
\resizebox{\columnwidth}{!}{
\begin{tabular}{c c c | c c c}
    \toprule
    \textbf{Device} & \textbf{Acc.} & \textbf{Matched/Total} 
    & \textbf{Device} & \textbf{Acc.} & \textbf{Matched/Total} \\
    \midrule
    \cellcolor{green!35}A & \cellcolor{green!35}88.9\% & \cellcolor{green!35}40/45
        & \cellcolor{green!35}M & \cellcolor{green!35}90.5\% & \cellcolor{green!35}19/21 \\
    \cellcolor{green!35}B & \cellcolor{green!35}100.0\% & \cellcolor{green!35}129/129
        & \cellcolor{green!35}N & \cellcolor{green!35}100.0\% & \cellcolor{green!35}27/27 \\
    \cellcolor{green!35}C & \cellcolor{green!35}100.0\% & \cellcolor{green!35}1236/1236
        & \cellcolor{orange!40}O & \cellcolor{orange!40}14.5\% & \cellcolor{orange!40}9/62 \\
    \cellcolor{green!35}D & \cellcolor{green!35}100.0\% & \cellcolor{green!35}6/6
        & \cellcolor{green!35}Q & \cellcolor{green!35}100.0\% & \cellcolor{green!35}98/98 \\
    \cellcolor{red!30}E & \cellcolor{red!30}0.0\% & \cellcolor{red!30}0/115
        & \cellcolor{green!35}R & \cellcolor{green!35}100.0\% & \cellcolor{green!35}287/287 \\
    \cellcolor{green!35}G & \cellcolor{green!35}100.0\% & \cellcolor{green!35}617/617
        & \cellcolor{green!35}S & \cellcolor{green!35}100.0\% & \cellcolor{green!35}64/64 \\
    \cellcolor{green!35}H & \cellcolor{green!35}100.0\% & \cellcolor{green!35}12/12
        & \cellcolor{yellow!40}T & \cellcolor{yellow!40}70.6\% & \cellcolor{yellow!40}12/17 \\
    \cellcolor{green!35}I & \cellcolor{green!35}100.0\% & \cellcolor{green!35}28/28
        & \cellcolor{red!30}U & \cellcolor{red!30}0.0\% & \cellcolor{red!30}0/5 \\
    \cellcolor{orange!40}J & \cellcolor{orange!40}44.2\% & \cellcolor{orange!40}73/165
        & \cellcolor{red!30}V & \cellcolor{red!30}0.0\% & \cellcolor{red!30}0/27 \\
    \cellcolor{green!35}K & \cellcolor{green!35}82.4\% & \cellcolor{green!35}28/34
        & \cellcolor{green!35}W & \cellcolor{green!35}100.0\% & \cellcolor{green!35}539/539 \\
    \cellcolor{red!30}L & \cellcolor{red!30}0.0\% & \cellcolor{red!30}0/54
        & \cellcolor{red!30}X & \cellcolor{red!30}0.0\% & \cellcolor{red!30}0/7 \\
    \bottomrule
\end{tabular}%
}
\end{table}

While setup S3.3 (Decomposed subfields, w/o IFAT) achieved a high global accuracy of 89.6\% using DBSCAN, this result can be misleading. Because devices do not transmit the same number of Probe Frames, this class imbalance has an impact on individual identification accuracy (Table~\ref{tab:individual_accuracies}). Devices with very few samples are frequently misclustered; for example, sparse transmitters like devices O and T are often absorbed by dominant devices like Q. Consequently, extreme minority classes (such as U, V, and X) exhibit 0\% individual accuracy. Due to their sparse data, these Probe Frames are easily absorbed into the closest dominant neighboring clusters. Devices L and V overlap and cluster with device B, while device U is absorbed by device W. 
Beyond sparse data, the clustering task is further complicated when a single device broadcasts two different \ac{HT} capabilities information values. This is seen with devices J and M, where the use of multiple signatures makes it nearly impossible to achieve 100\% individual accuracy. By splitting their Probe Frames across different configurations, their maximum accuracies were limited to 44.2\% and 90.5\%, respectively.

\subsection{Key findings and future directions}

The individual analysis revealed three primary challenges that limit obtaining higher accuracy:

\begin{itemize}
    \item \textbf{Low-frequency transmitters}: Devices that transmit very few bursts of Probe Frames do not generate sufficiently dense regions to be considered clusters.
    \item \textbf{Absence of \ac{HT} capabilities information}: Devices that do not broadcast \ac{HT} capabilities information provide less identifying information, forcing algorithms to rely on weaker secondary features or spatial proximity.
    \item \textbf{Multiple \ac{HT} capabilities information signatures}: Individual devices can alternate \ac{HT} capabilities information over time, making it difficult to cluster all their Probe Frames into a single device entity.
\end{itemize}

Future work could include extending this study to newer Wi-Fi standards such as \ac{VHT}, \ac{HE}, and \ac{EHT}. Additionally, temporal models such as \acf{LSTM} networks could be investigated to better capture user mobility and changes in device behavior in real-world scenarios.

\section{Conclusions} \label{sec:conclusion}

The main objective of this study was to evaluate the tracking capabilities of an eavesdropper to expose the vulnerabilities of Wi-Fi \ac{MAC} address randomization. Our findings demonstrate that the information inherent in Probe Requests is sufficient to bypass such a privacy measure using \acl{ML}-based fingerprinting. By evaluating static, temporal, and spatial features across three clustering algorithms, we found that decomposing the raw hexadecimal \ac{HT} capabilities information into 16-bit individual subfields is critical for accurate device separation. Furthermore, while \ac{IFAT} introduced noise for density-based algorithms, the integration of simulated spatial data (\ac{RSSI}) from a multi-sniffer architecture significantly improved identification. Overall, DBSCAN was shown to be the most robust model with a global accuracy of 89.6\% across 22 devices. These insights can serve as a reference for future Wi-Fi standards to enhance privacy protocols.

\section*{Acknowledgment}

This work was supported by the following projects: TRUE Wi-Fi PID2024-155470NB-I00 (MICIU/AEI/10,13039/501100011033/FEDER,UE), ICREA Academia 2024 (00077 AGAUR), MdM CEX2021-001195-M (MICIU/AEI/10.13039/501100011033), and by NextGenerationEU (Italian NRRP, Mission 4, Component 2, Investment 1.2, CUP F23C25000440006).


\bibliographystyle{IEEEtran}
\bibliography{references}

\end{document}